\documentclass[9pt,twocolumn,twoside]{osajnl}
\usepackage{nicefrac} 

\journal{ol} 

\setboolean{shortarticle}{true}

\title{Kilowatt-average-power compression of millijoule pulses in a gas-filled multi-pass cell}

\author[1,2,*]{Christian Grebing}
\author[1]{Michael M\"uller}
\author[1]{Joachim Buldt}
\author[1]{Henning Stark}
\author[1,2,3]{Jens Limpert}

\affil[1]{Institute of Applied Physics, Abbe Center of Photonics, Friedrich Schiller University Jena,
Albert-Einstein-Str. 6, 07745 Jena, Germany}
\affil[2]{Fraunhofer Institute for Applied Optics and Precision Engineering, Albert-Einstein-Str. 7, 07745 Jena, Germany}
\affil[3]{Helmholtz-Institute Jena, Fr\"obelstieg 3, 07743 Jena, Germany}

\affil[*]{Corresponding author: christian.grebing@uni-jena.de}

\doi{\url{https://doi.org/10.1364/OL.408998}}

\begin{abstract}
We demonstrate the reliable generation of 1-mJ, 31\hbox{-}fs pulses with an average power of 1\,kW by post-compression of 200-fs pulses from a coherently combined Yb:fiber laser system in an argon-filled Herriott-type multi-pass cell with an overall compression efficiency of 96\%. We also analyze the output beam, revealing essentially no spatio-spectral couplings or beam quality loss.
\end{abstract}

\setboolean{displaycopyright}{true}

\begin{document}

\maketitle

\textbf{Introduction.} Intense sub-100\,fs laser pulses can be used to drive unique secondary sources either by nonlinear frequency conversion into the Terahertz \cite{Buldt2019,Kramer2020} and X-ray regime 
\cite{Haedrich2016a,Schoenlein2019} or by acceleration of particles in a plasma \cite{Mangles2004,Faure2018}. Increasing the average power of the driving laser source leads to an increased average number of photons/particles of the secondary source and, thus, reduces acquisition times and improves signal-to-noise ratios in a plethora of applications. 

Ultrafast Ti:Sapphire laser systems have been the working horses in many scientific applications for many years. They readily provide sub-30\,fs pulses, but their average power is limited to only a few watts. In contrast, ultrafast Yb-based amplifiers are power-scalable into the kW regime \cite{Russbueldt2010,Nubbemeyer2017, Mueller2020}, however, their output pulse duration is typically not shorter than 200\,fs. 
Furthermore, highly efficient pulse compression in noble gas-filled compact multi-pass cells (MPC) has recently been utilized to compress the pulse duration 
\textcolor{black}{ of laser pulses in a broad energy range \cite{Milosevic2000,Ueffing2018, Kaumanns2018, Lavenu2018, Lavenu2019, Russbueldt2019, Balla2020, Kramer2020}. Similar to gas-filled hollow-core-fiber (HCF) post-compression setups \cite{Nagy2019, Jeong2018} the footprint of MPC setups scales with pulse energy leading to meter-scale assemblies in the multi-mJ range \cite{Ueffing2018, Kaumanns2018, Russbueldt2019, Balla2020, Kramer2020}. However, compared to HCF schemes the losses in an MPC setup can be much lower, which is advantageous for average power scaling.}

In this letter, we present the combination of a millijoule-energy high-average power Yb:fiber laser source with a highly efficient MPC compression stage demonstrating a close-to lossless peak power boost from 
\textcolor{black}{4.5\,GW to 24\,GW}. 
The experimental results represent the highest average power of sub-100\,fs pulses generated to date.

\begin{figure}[htbp]
\centering
\includegraphics[width=0.6\linewidth]{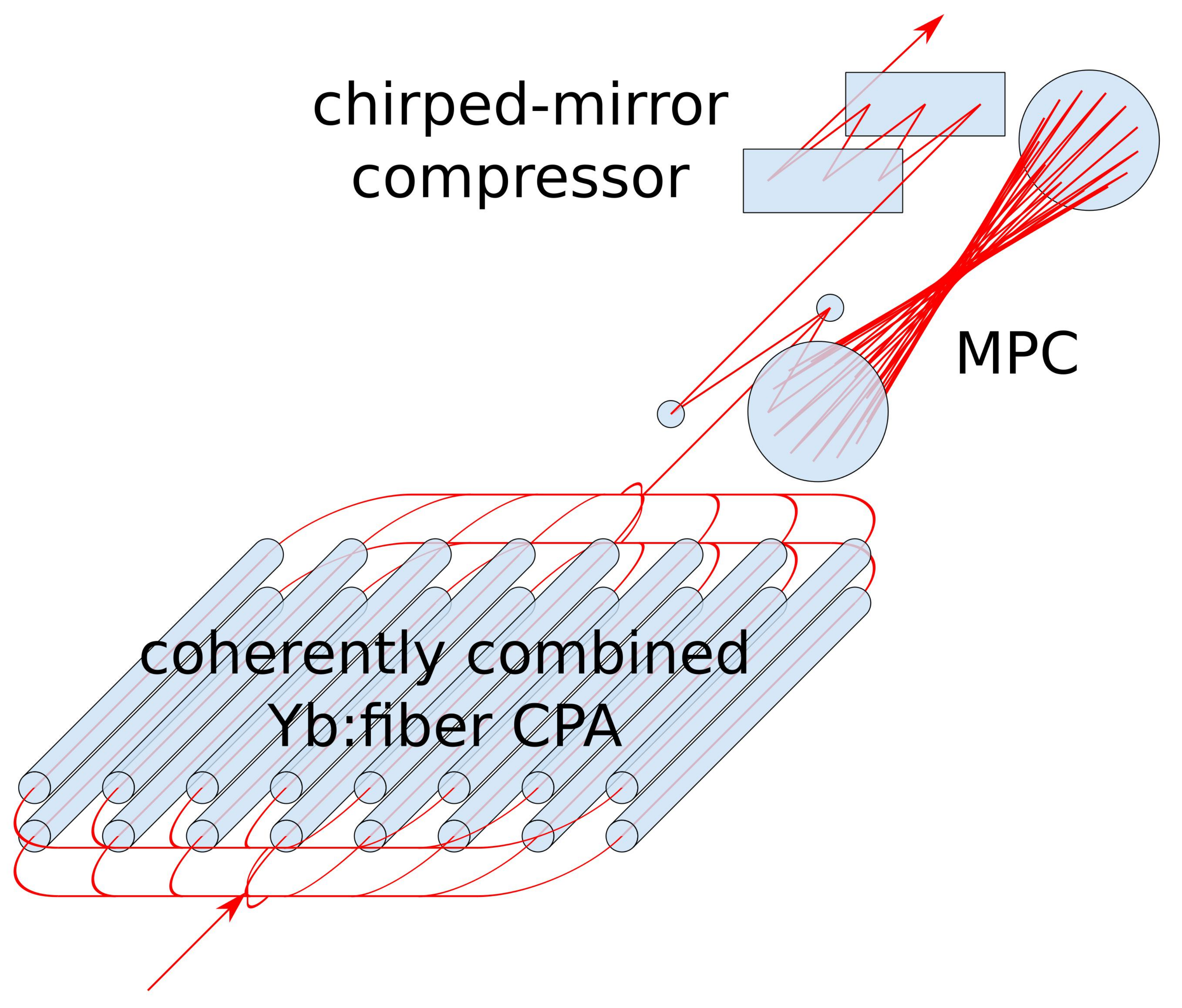}
\caption{Schematic of the experimental setup: Pulses generated from a high-power Yb:fiber chirped-pulse amplifier (CPA) system are fed into a nonlinear compression stage including spectral broadening in a multi-pass cell (MPC) and subsequent chirped-mirror compression.}
\label{fig:setup}
\end{figure}

\textbf{Setup.} The optical setup is depicted in Fig.\,\ref{fig:setup}. The laser source for the pulse compression is a high-power Yb:fiber chirped-pulse amplifier system. The main amplifier stage coherently combines the output of 16 fiber amplifier channels operated in parallel \cite{Mueller2018}. The system provides pulses with $\sim$1-mJ pulse energy and a pulse duration $\tau^{\rm FWHM}_{\rm in}$ of 200\,fs at a repetition rate of 1\,MHz resulting in an average power of more than 1\,kW.
\textcolor{black}{The output pulse duration is optimized close to its transform limit using the multiphoton intrapulse interference phase scan technique \cite{Lozovoy2004} and can be fine-tuned with a tunable dispersive filter.}

The MPC itself consists of two concave mirrors with a radius-of-curvature of 0.6\,m and 75\,mm diameter placed in a 1.5\,m long and 0.4\,m wide low-pressure chamber with two 3\,mm thick fused silica windows for in- and out-coupling. The MPC mirror distance is adjusted to 1183\,mm resulting in 26 focal passes $n_{\rm foci}$. Thus, the MPC forms a near concentric cavity operated close to its stability edge. The calculated Gaussian eigenmode for a wavelength of 1030\,nm exhibits a 0.30\,mm 1/$e^2$-diameter at the focal position and a 2.5\,mm 1/$e^2$-diameter on the mirror surface. The dimensions of the MPC are matched to the laser parameters taking two limitations into account: in-focus ionization and optical damage of the MPC mirrors. The mirror distance is chosen such that after the 26 focal passes the ray hits again the entrance point of the MPC (re-entrance condition \cite{Herriott1964}), which constitutes a close-to q-preserving configuration \cite{Sennaroglu2004}. This way, only one plane scraper mirror is needed, which enables in- as well as out-coupling to the MPC. The outgoing beam returns at a slightly different angle compared to the incoming beam, which enables separation of both. The beam is mode matched to the eigenmode of the MPC using a 4-m radius-of-curvature concave mirror and an identical one is used for re-collimation behind the MPC. 
\textcolor{black}{The cell is operated below the ionization threshold of the noble gas filling, which otherwise introduces an additional loss channel and beam quality degradation. The chosen configuration exhibits a focal peak intensity of about 10 TW/cm$^2$ , which calls for argon, neon or helium as a filling gas \cite{Vozzi2004}. Argon was chosen as it exhibits a reasonable nonlinearity of $n_2=1\times10^{-23}$\,m$^2$/W at atmospheric pressure to trigger self-phase modulation combined with a low dispersion.}
Hence, the low-pressure chamber is filled with argon at a pressure of 700\,mbar introducing a moderate nonlinear phase shift per single focal pass of $\phi^{nl}_{\rm focus} \sim 0.5$\,rad (averaged over the transverse beam profile) \cite{Hanna2017}. 
At all points during the spectral broadening the residual propagation path in the chamber $L_{\rm prop}$ is much smaller than the dispersion length $L_d$ \cite{Agrawal2013} (after propagation of $L_d$ a Gaussian pulse's duration increased by $\sqrt{2}$), e.g. for the input pulse $L_d (\sim 2\,{\rm km})\gg L_{\rm prop}(\sim 35\,{\rm m})$. Hence, the propagation can be considered dispersion-free. So, the total average phase shift of the MPC passage is expected to be $\phi^{nl}_{\rm total} = n_{\rm foci}\times \phi^{nl}_{\rm focus}$ leading to a theoretical compression factor of $\nicefrac{\tau^{\rm FWHM}_{\rm in}}{\tau^{\rm FWHM}_{\rm out}}=\phi^{nl}_{\rm total}/\sqrt{e}=8.1$ under the assumption of a Gaussian pulse shape \cite{Kaumanns2020}.
The chirp after the MPC is removed with a chirped-mirror compressor comprising
six bounces off chirped mirrors each having a group delay dispersion (GDD) of -350\,fs$^2$. The total GDD is -1923\,fs$^2$ after fine-tuning the compression with inserted glass plates. The compressor is placed inside the low-pressure chamber to ensure minimum nonlinear distortion during compression.

\begin{figure}[htbp]
\centering
\includegraphics[width=\linewidth]{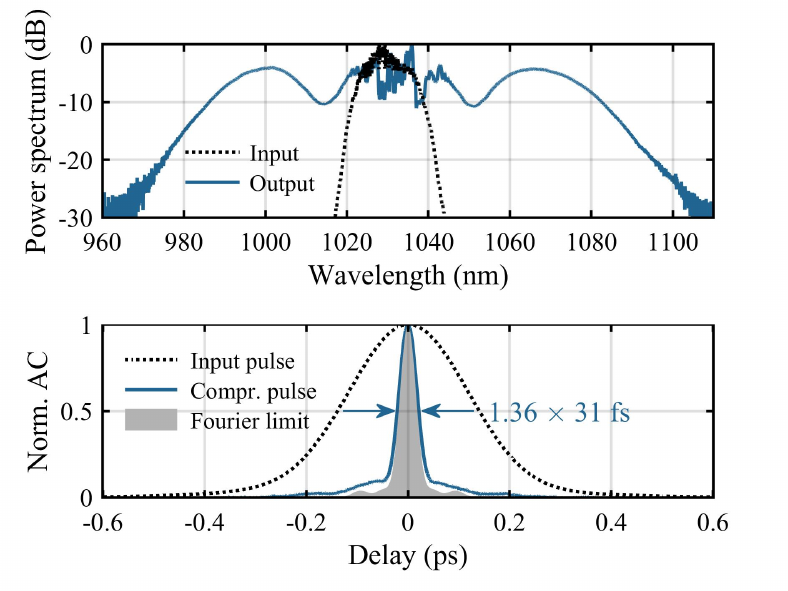}
\caption{\textcolor{black}{Upper panel: Measured input (dotted line) and compressed output spectrum (blue line). Lower panel: Comparison of normalized autocorrelation traces of the compressed (blue line) pulse, the input pulse (dotted line), and the Fourier-limited pulse (gray area). The latter is derived from the output spectrum in the upper panel.}}
\label{fig:AC}
\end{figure}

\textbf{Results.} The measured broadened spectrum behind the compression chamber is shown in the upper panel of Fig.\,\ref{fig:AC} exhibiting a 20-dB-width of $\sim$120\,nm, which corresponds to a Fourier-limited pulse duration of 27\,fs and leads to a maximum compression factor of 7.4, which is close to the one estimated above. The spectrum reveals symmetric side-wings indicating self-phase-modulation dominated broadening. The compressed pulse duration is inferred from a second-order intensity autocorrelation measurement to 31\,fs and is close to the computed Fourier-limited autocorrelation trace (see Fig.\,\ref{fig:AC}, lower panel). The deconvolution factor was calculated from the measured spectrum. 
\textcolor{black}{From the pedestal area of the autocorrelation trace we estimate a fraction of 80\% of the energy in the main pulse, which is a typical value for SPM-based pulse post-compression.}
The compressed pulse duration is limited by the bandwidth of the used broadband dielectric mirrors (for the wavelength range $970-1130$\,nm: $R>99.95\%$, ${\rm |GDD|}<50\,{\rm fs}^2$). 

\begin{figure}[htbp]
\centering\includegraphics{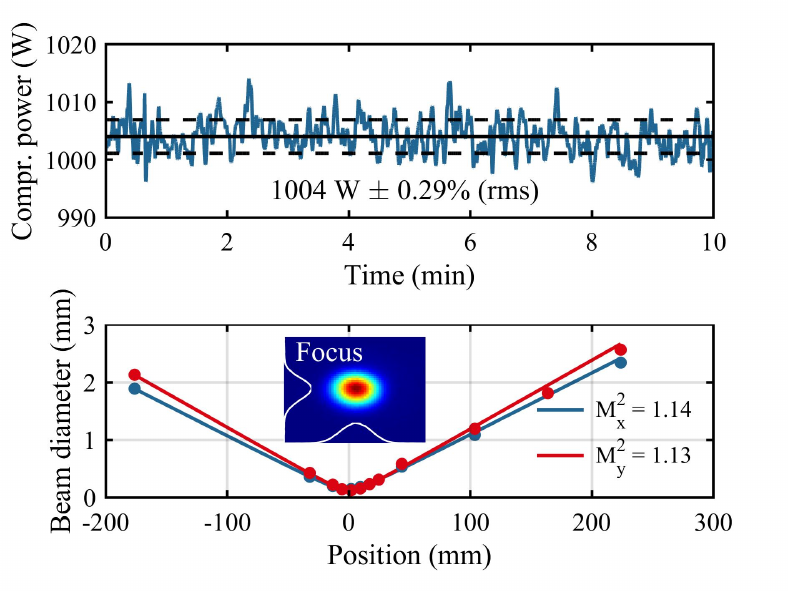}
\caption{Upper panel: Long-term measurement of the compressed output power \textcolor{black}{employing a thermopile sensor with a sample frequency of 10\,Hz. } 
Lower panel: Measurement of the beam propagation factor of the compressed output beam along the transverse x- (blue) and y-axis (red). Inset: snapshot of the focal intensity distribution.}
\label{fig:M2pow}
\end{figure}

A long-term measurement of the compressed output power is depicted in the upper panel of Fig.\,\ref{fig:M2pow} demonstrating an excellent power stability of $1004\,{\rm W} \pm 0.29\%$, which implies negligible thermal drifts even at highest average power. The input power was 1045 W 
\textcolor{black}{exhibiting the same stability} 
resulting in a combined transmission of MPC and chirped-mirror compressor of 96\%. The transmission is close to what is expected from $\sim$40 bounces of highly reflective broadband dielectric mirrors and is independent of the input power. This confirms operation of the MPC below the ionization threshold of argon.
To the best of our knowledge, the MPC output represents the highest average power of sub-100\,fs pulses ever demonstrated. Recently, it was shown that an even higher nonlinear phase shift per focus does not deteriorate the MPC performance \cite{Kramer2020,Daher2020a}. Hence, further reduction of MPC round-trips seems feasible, which could in turn enable an even higher transmission.

The quality of the compressed output beam was characterized in terms of the beam propagation factor $M^2$ and of the spatio-spectral homogeneity. The lower panel of Fig.\,\ref{fig:M2pow} shows the beam diameter measurements along x- and y-axis for a focus transition of an attenuated fraction of the output beam. The measurements were analyzed yielding an $M^2$ of smaller than 1.15 for both axes, basically resembling the $M^2$ of the input beam. 

\begin{figure}[htbp]
\centering\includegraphics{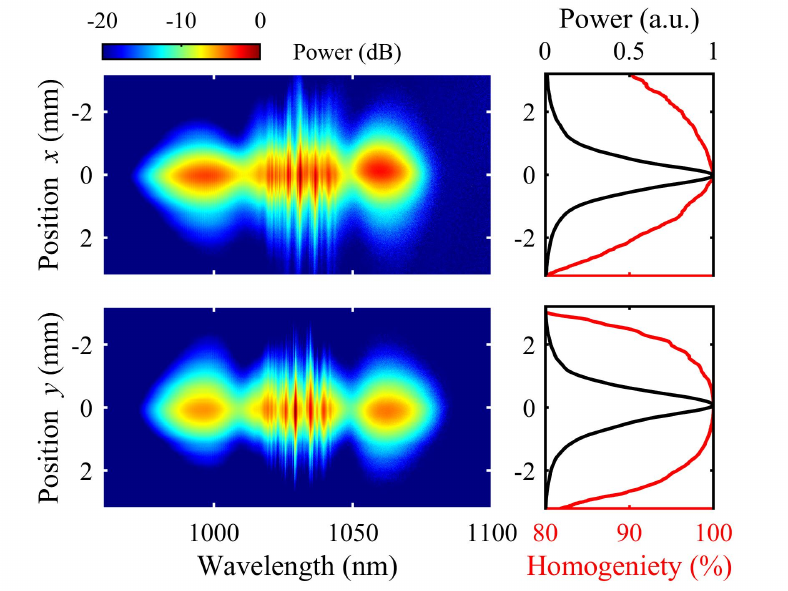}
\caption{Left: Spatio-spectral map along the transverse x- (upper panel) and y-direction (lower panel) of the output beam. Right: spatio-spectral map integrated along the wavelength axis (black line) and corresponding homogeneity computed following the definition in \cite{Weitenberg2017}.}
\label{fig:Hom}
\end{figure}

Spectral broadening by self-phase modulation can be introduced by focusing a laser beam in a nonlinear medium. However, the Kerr nonlinearity introduces a spatial variation of the nonlinear phase and, hence, of the generated spectral components depending on the beam profile \cite{Milosevic2000}. Thus, a spatio-spectral inhomogeneity occurs after a single focal pass with a significant nonlinear phase shift. This leads to beam quality degradation and low pulse quality \cite{Seidel2016}. For multiple foci, e.g. as achieved in a MPC, the spatio-spectral coupling is significantly reduced due to free-space mode coupling \cite{Milosevic2000}. 
In order to study this homogenization effect we set up an imaging spectrograph composed of a 1200\,lines/mm grating, a 50-mm focal length cylindrical lens, a Si-CCD  camera ($1920\times1200$\,pixels, pixel size: 5.86\,\textmu m), and an input telescope to match the input beam diameter to the CCD chip size. 
\textcolor{black}{Thus, the horizontal dimension on the chip corresponds to the optical spectrum and the vertical dimension to the radial coordinate}
The corresponding spatio-spectral maps along x- and y-axis are depicted in the left panel of Fig.\,\ref{fig:Hom}. From that map we analyze the homogeneity as defined in \cite{Weitenberg2017} as a figure-of-merit that quantifies the overlap of off-axis spectra with the on-axis spectrum. The homogeneity computed for the measured maps are shown in the right panel of Fig.\,\ref{fig:Hom} in red. The homogeneity is >80\% even at the outer edges of the beam profile. For comparison, the homogeneity was weighted with the beam profile, which was computed by integrating the spatio-spectral map over the wavelength axis (black lines in the right panel of Fig.\,\ref{fig:Hom}) yielding an effective homogeneity of 98.7\% (x-axis) and 99.4\% (y-axis), which resembles the homogeneity of the uncompressed beam.

Therefore, the highly efficient compression to the 30-fs range does not cause any measurable distortion to the compressed beam neither in spatial beam quality nor in spatio-spectral homogeneity. The setup realizes a close-to lossless compression at highest average power.

\textbf{Conclusion.} We have demonstrated the scaling of 1-mJ, 31-fs pulses to a record average power of 1\,kW. The source comprises a coherently-combined high-power Yb:fiber laser system and a subsequent close-to lossless nonlinear compression stage. The compact-footprint nonlinear compressor utilizes gas-filled MPC spectral broadening and a chirped-mirror compression. The nonlinear compression stage features a transmission of 96\%, an excellent power stability, an unperturbed beam quality, and negligible spatio-spectral couplings. Accordingly, the presented system is a unique driver for high-average power sources in the THz and X-ray regime. In terms of pulse duration, the compression is limited by the bandwidth of available dielectric mirrors. MPC driven compression to well below the 30-fs range requires low-dispersion mirrors with substantially larger bandwidth, which so far is attainable only with metallic coatings. However, metallic coatings exhibit a lower reflectivity compared to dielectric coatings leading to a reduced transmission \cite{Balla2020}. Increasing the pulse energy leads to ionization induced losses and/or mirror damage. Thus, going to higher pulse energies requires a larger MPC \cite{Kaumanns2018, Kaumanns2020, Kramer2020} or other mitigation strategies, e.g. longer input pulses and different gases.

\medskip
\noindent\textbf{Acknowledgments.} This project has received funding from the European Research Council (ERC) under the  European Union’s Horizon 2020 Research and Innovation Programme (Grant Agreement No.835306), the Fraunhofer Cluster of Excellence “AdvancedPhoton Sources,” and the Bundesministerium für Bildung und Forschung (BMBF) (01DR20009A).
%

\medskip
\noindent\textbf{Disclosures.} The authors declare no conflicts of interest.

\bibliography{paper_db_cg}



\ifthenelse{\equal{\journalref}{aop}}{%
\section*{Author Biographies}
\begingroup
\setlength\intextsep{0pt}
\begin{minipage}[t][6.3cm][t]{1.0\textwidth} 
  \begin{wrapfigure}{L}{0.25\textwidth}
    \includegraphics[width=0.25\textwidth]{john_smith.eps}
  \end{wrapfigure}
  \noindent
  {\bfseries John Smith} received his BSc (Mathematics) in 2000 from The University of Maryland. His research interests include lasers and optics.
\end{minipage}
\begin{minipage}{1.0\textwidth}
  \begin{wrapfigure}{L}{0.25\textwidth}
    \includegraphics[width=0.25\textwidth]{alice_smith.eps}
  \end{wrapfigure}
  \noindent
  {\bfseries Alice Smith} also received her BSc (Mathematics) in 2000 from The University of Maryland. Her research interests also include lasers and optics.
\end{minipage}
\endgroup
}{}

\end{document}